\documentclass[twocolumn,aps,pra,showpacs,superscriptaddress,floatfix]{revtex4}
\usepackage{graphicx}
\usepackage{times}
\usepackage{nicefrac}
\usepackage{amsmath}
\usepackage{amsfonts}
\usepackage{amssymb}
\usepackage{amsthm}
\usepackage{epsf}
\usepackage{bm}
\usepackage{bbm}
\usepackage{times}

\usepackage{dcolumn}

\newcolumntype{.}{D{x}{}{-1}}

\begin{document}

\newcommand{\half}{\frac12}
\newcommand{\vare}{\varepsilon}
\newcommand{\eps}{\epsilon}
\newcommand{\pr}{^{\prime}}
\newcommand{\ppr}{^{\prime\prime}}
\newcommand{\pp}{{p^{\prime}}}
\newcommand{\ppp}{{p^{\prime\prime}}}
\newcommand{\hp}{\hat{\bfp}}
\newcommand{\hr}{\hat{\bfr}}
\newcommand{\hx}{\hat{\bfx}}
\newcommand{\hpp}{\hat{\bfpp}}
\newcommand{\hq}{\hat{\bfq}}
\newcommand{\rqq}{{\rm q}}
\newcommand{\bfk}{{\bm{k}}}
\newcommand{\bfp}{{\bm{p}}}
\newcommand{\bfq}{{\bm{q}}}
\newcommand{\bfr}{{\bm{r}}}
\newcommand{\bfx}{{\bm{x}}}
\newcommand{\bfy}{{\bm{y}}}
\newcommand{\bfz}{{\bm{z}}}
\newcommand{\bfpp}{{\bm{\pp}}}
\newcommand{\bfppp}{{\bm{\ppp}}}
\newcommand{\balpha}{\bm{\alpha}}
\newcommand{\bvare}{\bm{\vare}}
\newcommand{\bgamma}{\bm{\gamma}}
\newcommand{\bGamma}{\bm{\Gamma}}
\newcommand{\bLambda}{\bm{\Lambda}}
\newcommand{\bmu}{\bm{\mu}}
\newcommand{\bnabla}{\bm{\nabla}}
\newcommand{\bvarrho}{\bm{\varrho}}
\newcommand{\bsigma}{\bm{\sigma}}
\newcommand{\bTheta}{\bm{\Theta}}
\newcommand{\bphi}{\bm{\phi}}
\newcommand{\bomega}{\bm{\omega}}
\newcommand{\intzo}{\int_0^1}
\newcommand{\intinf}{\int^{\infty}_{-\infty}}
\newcommand{\lbr}{\langle}
\newcommand{\rbr}{\rangle}
\newcommand{\ThreeJ}[6]{
        \left(
        \begin{array}{ccc}
        #1  & #2  & #3 \\
        #4  & #5  & #6 \\
        \end{array}
        \right)
        }
\newcommand{\SixJ}[6]{
        \left\{
        \begin{array}{ccc}
        #1  & #2  & #3 \\
        #4  & #5  & #6 \\
        \end{array}
        \right\}
        }
\newcommand{\NineJ}[9]{
        \left\{
        \begin{array}{ccc}
        #1  & #2  & #3 \\
        #4  & #5  & #6 \\
        #7  & #8  & #9 \\
        \end{array}
        \right\}
        }
\newcommand{\Dmatrix}[4]{
        \left(
        \begin{array}{cc}
        #1  & #2   \\
        #3  & #4   \\
        \end{array}
        \right)
        }
\newcommand{\Dcase}[4]{
        \left\{
        \begin{array}{cl}
        #1  & #2   \\
        #3  & #4   \\
        \end{array}
        \right.
        }
\newcommand{\cross}[1]{#1\!\!\!/}

\newcommand{\Za}{{Z \alpha}}
\newcommand{\im}{{\rm i}}

\newcommand{\hfs}{{\em hfs }}

%%%%%%%%%%%%%%%%%%%%%%%%%%%%%%%%%%%%%%%%%%%%%%%%%%%%%%%%%%
\title{Improved all-order
results for the one-loop QED correction to the hyperfine structure
in light H-like atoms}

\author{V. A. Yerokhin}
\affiliation{Department of Physics, St.~Petersburg State
University, Oulianovskaya 1, Petrodvorets, St.~Petersburg 198504,
Russia}

\affiliation{Center for Advanced Studies, St.~Petersburg State
Polytechnical University, Polytekhnicheskaya 29, St.~Petersburg
195251, Russia}

\author{A. N. Artemyev}
\affiliation{Department of Physics, St.~Petersburg State
University, Oulianovskaya 1, Petrodvorets, St.~Petersburg 198504,
Russia}

\author{V. M. Shabaev}
\affiliation{Department of Physics, St.~Petersburg State
University, Oulianovskaya 1, Petrodvorets, St.~Petersburg 198504,
Russia}

\affiliation{Max-Planck-Institut f\"ur Physik komplexer Systeme,
N\"othnitzer Str. 38, D-01187 Dresden, Germany}

\author{G. Plunien}
\affiliation{Institut f\"ur Theoretische Physik, TU Dresden,
Mommsenstrasse 13, D-01062 Dresden, Germany}

\begin{abstract}

A calculation of the one-loop self-energy and vacuum-polarization
corrections to the hyperfine splitting of the $1s$ and $2s$ states
in light H-like ions is carried out to all orders in the parameter
$\Za$. Using the known values for the $\Za$-expansion
coefficients, the numerical data obtained are extrapolated from
$Z=5$ and higher to $Z=0$, 1, and 2, with the resulting accuracy
being significantly better than in previous evaluations. Our
calculation shifts the theoretical value of the normalized
difference of the $1s$ and $2s$ hyperfine-structure intervals in
$^3{\rm He}^+$ by $0.056$~kHz and improves its accuracy.

\end{abstract}


\maketitle

%%%%%%%%%%%%%%%%%%%%%%%%%%%
\section*{Introduction}

Hyperfine splitting of the ground state in light H-like systems,
such as hydrogen, deuterium, tritium, and helium-3 ion, has long
been known experimentally with extremely high precision. The
present-day theory of the ground-state hyperfine structure ({\em
hfs}) is still far behind the experiment, due to a relatively
large contribution of the nuclear-structure effects, which cannot
be accurately calculated at present. One of the possibilities to
overcome this difficulty \cite{sternheim:63} is to study the
normalized difference $\Delta_{21} = 8\, \nu_{2s}-\nu_{1s}$, where
$\nu_{1s}$ and $\nu_{2s}$ are the $1s$ and the $2s$ \hfs interval,
respectively. A large class of corrections to $\nu_{1s}$ and
$\nu_{2s}$ (among them, all lowest-order nuclear effects) are
proportional to the nonrelativistic electron density at the
position of the nucleus ($r=0$) and, therefore, do not contribute
to the difference $\Delta_{21}$. Consequently, the theoretical
study of this difference can be performed up to a much higher
accuracy than that of $\nu_{1s}$ and $\nu_{2s}$ separately.

The experimental value of the difference $\Delta_{21}$ is obtained
by combining results of two independent measurements of $\nu_{1s}$
and $\nu_{2s}$ and is known less precisely than $\nu_{1s}$. The
best accuracy is obtained for the helium-3 ion in a combination of
two relatively old results \cite{schluessler:69,prior:77},
\begin{equation}\label{1}
    \Delta_{21}(^3{\rm He}^+) = 1\,189.979\,(71)\,\,{\rm kHz}\,.
\end{equation}
Recent progress was achieved in the measurement of $\nu_{2s}$
in hydrogen \cite{kolachevsky:04:prl} and
deuterium \cite{kolachevsky:04:pra}, which significantly improved the
corresponding experimental values for the difference $\Delta_{21}$.

Theory of \hfs and, specifically, of the difference $\Delta_{21}$
in light H-like atoms has recently been examined in detail in
Ref.~\cite{karshenboim:02:epjd}. It is demonstrated that one of
the major uncertainties in the theoretical prediction of
$\Delta_{21}(^3{\rm He}^+)$ stems from the one-loop self-energy
correction. The self-energy correction is also responsible for a
significant part of the theoretical uncertainty for the
ground-state hyperfine splitting in muonium \cite{mohr:05:rmp}.

The goal of the present investigation is to improve the numerical
accuracy of the one-loop QED correction for the $1s$ and $2s$
states in light H-like atoms. Our consideration will be carried
out to all orders in the parameter $\Za$ ($Z$ is the nuclear
charge number and $\alpha$ is the fine-structure constant).
All-order calculations of the self-energy \hfs correction in
H-like ions have been previously performed by numerous authors
\cite{persson:96:hfs,yerokhin:96:pisma,blundell:97:pra,yerokhin:97:eprint,%
blundell:97:prl,sunnergren:98:pra,yerokhin:01:hfs,sapirstein:01:hfs}.
Different evaluations are generally in good agreement (except for
the first two calculations, see a discussion in
Ref.~\cite{yerokhin:01:hfs}). For high- and middle-$Z$ ions,
results for this correction can be presently considered as well
established at the level of the experimental interest. In the
low-$Z$ region, however, the experimental accuracy is much higher
and technical problems encountered in all-order calculations are
more demanding than for higher-$Z$ ions. It would be clearly
preferable to perform a direct all-order calculation for $Z=1$ and
2 with an accuracy significantly higher than the one obtained from
the $\Za$ expansion, as it was done for the Lamb shift
\cite{jentschura:99:prl}. However, such a project has not been
realized yet. Blundell {\em et al.} \cite{blundell:97:prl}
obtained the higher-order (in $\Za$) contribution to the $1s$
self-energy correction for $Z=1$ by extrapolating their numerical
results for higher $Z$. Similar procedure was employed in the
investigation by two of us \cite{yerokhin:01:hfs} for the
self-energy correction to the $1s$ and $2s$ \hfs intervals for
$Z=1$ and 2.

The vacuum-polarization \hfs correction was evaluated to all
orders in $\Za$ in
Refs.~\cite{persson:96:hfs,shabaev:97:pra,shabaev:98:hfs} (without
the magnetic-loop Wichmann-Kroll correction) and in
Refs.~\cite{sunnergren:98:pra,artemyev:01} (complete
calculations). However, the above studies were mainly concerned
with high- and middle-$Z$ ions, so that little information was
provided about the behavior of the Wichmann-Kroll part of the
vacuum-polarization correction in the low-$Z$ region.

In the present work, we perform a calculation of the one-loop
self-energy and vacuum-polarization corrections to the $1s$ and
$2s$ \hfs intervals in H-like ions. The paper is organized as
follows. In Sec.~\ref{sec:se} we evaluate the self-energy
correction by employing the additional-subtraction scheme
\cite{yerokhin:05:se} that improves the convergence properties of
the resulting partial-wave expansion. In this way, we
significantly increase the accuracy of the numerical results for
$Z \ge 5$ as compared to the previous evaluations. The
vacuum-polarization correction is evaluated in Sec.~\ref{sec:vp}.
The higher-order one-loop QED contribution is inferred from our
all-order results  in Sec.~\ref{sec:ho} by subtracting the known
terms of the $\Za$ expansion. Finally, we discuss the experimental
consequences of our calculation.

Relativistic units ($\hbar = c = m = 1$) are used throughout the paper.

%%%%%%%%%%%%%%%%%%%%%%%%%%%%%%%%%%%%%%%%%%%%%%%
%
%
%
%%%%%%%%%%%%%%%%%%%%%%%%%%%%%%%%%%%%%%%%%%%%%%%
\section{Self-energy correction}
\label{sec:se}

In this section we describe the evaluation of the self-energy \hfs correction
without any expansion in $\Za$. We start with general
formulas for the self-energy correction in the presence of an
additional perturbing potential $\delta V$. To the first order in
$\delta V$, the self-energy correction is given by the sum of the
{\em irreducible}, the {\em reducible}, and the {\em vertex}
correction \cite{shabaev:02:rep},
\begin{equation}\label{4a}
    \Delta E_{\rm SE} = \Delta E_{\rm ir}+\Delta E_{\rm red}+\Delta E_{\rm
    ver}\,.
\end{equation}
The irreducible part arises through a perturbation of the wave
function,
\begin{equation}\label{5}
\Delta E_{\rm ir} = \lbr \delta a|\gamma^0
    \widetilde{\Sigma}(\vare_a) |a\rbr
              + \lbr a|\gamma^0 \widetilde{\Sigma}(\vare_a) |\delta a\rbr\,,
\end{equation}
where $\widetilde{\Sigma} = \Sigma -\delta m$, $\delta m$ is the
one-loop mass counterterm, $\Sigma$ is the one-loop self-energy
function,
\begin{eqnarray} \label{6}
 \Sigma(\vare,\bfx_1,\bfx_2) &=&
2\,\im\alpha\,\gamma^0 \intinf d\omega\,
      \alpha_{\mu}\,
 \nonumber \\ && \times
         G(\vare-\omega,\bfx_1,\bfx_2)\, \alpha_{\nu}\,
    D^{\mu\nu}(\omega,\bfx_{12})\,
         \,,
\end{eqnarray}
$G$ is the Dirac Coulomb Green function $G(\vare) = [\vare-{\cal
H}(1-\im 0)]^{-1}$, $\cal H$ is the Dirac Coulomb Hamiltonian,
$D^{\mu\nu}$ is the photon propagator, ${\alpha}^{\mu} = (1,
\balpha)$, and $\bfx_{12} = \bfx_1-\bfx_2$. The perturbed wave
function is given by
\begin{equation}\label{7}
|\delta a\rbr = \sum_n^{\vare_n \ne \vare_a} \frac{|n\rbr
        \lbr n|\delta V|a\rbr}{\vare_a-\vare_n} \,.
\end{equation}
The reducible part can be considered as a correction due to the
first-order perturbation of the binding energy,
\begin{equation}\label{8}
    \Delta E_{\rm red} = \delta \vare_a\, \lbr a| \gamma^0 \left.
    \frac{\partial}{\partial \vare} \widetilde{\Sigma}(\vare) \right|_{\vare = \vare_a} |
            a\rbr\,,
\end{equation}
where $\delta \vare_a = \lbr a|\delta V|a\rbr$. The vertex part is
given by
\begin{align} \label{9}
&\Delta E_{\rm ver} = \ \frac{\im}{2\pi}\intinf d\omega\,
  \nonumber \\
& \times   \sum_{n_1n_2}
    \frac{\lbr n_1|\delta V|n_2\rbr\,
            \lbr an_2| I(\omega)|n_1a\rbr}
                   {[\vare_a-\omega-\vare_{n_1}(1-\im 0)]
                   [\vare_a-\omega-\vare_{n_2}(1-\im 0)]}\,,
\end{align}
where $I(\omega) = e^2\,
\alpha_{\mu}\alpha_{\nu}\,D^{\mu\nu}(\omega)$.

The self-energy correction to \hfs is given by the above formulas,
in which we should assume the perturbing potential to have the
form of the Fermi-Breit interaction (the nuclear magnetic moment
is denoted by $\bmu$),
\begin{equation}\label{10a}
    \delta V \to V_{hfs}(\bfr) =
    \frac{|e|}{4\pi}\,\frac{\balpha\cdot[\bmu\times\bfr]}{r^3}\,,
\end{equation}
and the initial-state wave function $|a\rbr$ to be the wave
function of the coupled system (electron$+$nucleus),
\begin{equation}\label{10}
    |a \rbr \to  |FM_FIj\rbr =  \sum_{M_Im_a} C^{FM_F}_{IM_I j_a
        m_a}\, |IM_I\rbr \,|j_a m_a\rbr\,,
\end{equation}
where $|IM_I\rbr$ denotes the nuclear wave function, $|j_a
m_a\rbr$ is the electron wave function, $F$ is the total momentum
of the atom, and $M_F$ is its projection. Radial integrations over
the nuclear coordinates can easily be performed already in the
general expressions. One can show that formulas
(\ref{5})-(\ref{9}) yield corrections to \hfs if we employ the
perturbing interaction in the form
\begin{equation}\label{11}
    \delta V(\bfr) = \frac{E_F}{4/3(\Za)^3}\, \frac{[\bfr \times
            \balpha]_z}{r^3}\,,
\end{equation}
where $E_F$ is the nonrelativistic Fermi energy, and consider the
initial-state wave function to be the electron wave function with
the moment projection $m_a = 1/2$,
\begin{equation}\label{12}
    |a\rbr = |j_a\,1/2\rbr\,.
\end{equation}

%%%%%%%%%%%%%%%%%%%%%%%%%%%%%%%%%%%%%%%%%%%%%%%
\subsection{Irreducible part}

As follows from Eq.~(\ref{5}), evaluation of the irreducible part
of the self-energy \hfs correction implies a calculation of a
non-diagonal matrix element of the self-energy function and,
therefore, is very similar to the evaluation of the first-order
self-energy correction to the Lamb shift. Since our present
approach to this problem is somewhat different from the standard
potential-expansion method, we now give a short description of the
scheme used for the evaluation of a self-energy matrix element.

Ultraviolet divergencies in the self-energy function (\ref{6}) are
traditionally isolated by separating the first two terms in the
expansion of the bound-electron propagator $G$ in terms of the
binding potential $V$,
\begin{eqnarray} \label{13}
G(E,\bfx_1,\bfx_2) &=&
G^{(0)}(E,\bfx_1,\bfx_2)+G^{(1)}(E,\bfx_1,\bfx_2)
  \nonumber \\ &&
                         +G^{(2+)}(E,\bfx_1,\bfx_2)\,,
\end{eqnarray}
where $G^{(0)} = [\omega-{\cal H}_0(1-\im 0)]^{-1}$ is the free
Dirac Green function, $G^{(1)}$ is the first-order expansion term
\begin{equation} \label{14}
G^{(1)}(E,\bfx_1,\bfx_2) = \int d\bfz\, G^{(0)}(E,\bfx_1,\bfz)\,
             V(z)\, G^{(0)}(E,\bfz,\bfx_2)\,,
\end{equation} and $G^{(2+)}$ is the remainder. Representing $G$
in the form (\ref{13}) leads to splitting the matrix element of
the self-energy function into the zero-potential, one-potential,
and many-potential parts (see Refs.
\cite{snyderman:91,blundell:92,yerokhin:99:pra} for details),
\begin{equation}\label{15}
\lbr a |\gamma^0 \widetilde{\Sigma}(\vare_a)|a\rbr
 = \Delta E_{\rm zero}+
    \Delta E_{\rm one}+
     \Delta E_{\rm \, many}\,,
\end{equation}
with the mass-counterterm part naturally ascribed to the
zero-potential term.

Modifications of the standard potential-expansion approach
introduced in our previous investigation \cite{yerokhin:05:se}
concern the many-potential term, which is given by
\begin{align} \label{16}
\Delta E_{\rm \, many} =&\  2\,\im\alpha \int_{C_{LH}} d\omega
\int d\bfx_1\,
        d \bfx_2\,
        D^{\mu \nu}(\omega, \bfx_{12})
  \nonumber \\ & \times
        \psi^{\dag}_a(\bfx_1)\, \alpha_{\mu}\,
        G^{(2+)}(\vare_a-\omega,\bfx_1,\bfx_2)\, \alpha_{\nu}\,
        \psi_a(\bfx_2)\,
        \,.
\end{align}
The integration contour $C_{LH}$ consists of two parts, the
low-energy ($C_L$) and the high-energy ($C_H$) one. The low-energy
part extends from $\vare_0-\im 0$ to $-\im 0$ on the lower bank of
the branch cut of the photon propagator and from $\im 0$ to
$\vare_0+\im 0$ on the upper bank of the cut. In order to avoid
the appearance of poles of the electron propagator near the
integration contour, each part of $C_L$ is bent into the complex
plane if the calculation is performed for an excited state. The
high-energy part of the contour is $C_H =
(\vare_0-\im\infty,\vare_0-\im 0]+[\vare_0+\im
0,\vare_0+\im\infty)$. The parameter $\vare_0$ of the contour is
chosen arbitrary from the interval $\vare_0 \in
(\vare_a-\vare_{1s},1+\vare_a)$, where $\vare_{1s}$ is the
ground-state energy.

The function $G^{(2+)} = G-G^{(0)}-G^{(1)}$ that enters
Eq.~(\ref{16}) is not known in its closed form at present and,
consequently, an evaluation of the many-potential term has to be
performed by expanding $G$ (and, therefore, $G^{(2+)}$) into
eigenfunctions of the Dirac angular momentum with the eigenvalue
$\kappa$. This expansion will be referred to as the partial-wave
expansion in the following. The convergence properties of this
expansion are of crucial importance for the numerical evaluation
of the self-energy correction. In our previous investigation
\cite{yerokhin:05:se}, it was demonstrated that the convergence
rate of the partial-wave expansion could be significantly enhanced
by separating from $\Delta E_{\rm \, many}$ a part that is
calculated in a closed form.

It was shown that in the region $\bfx_1 \approx \bfx_2$ the
function $G^{(2+)}$ could be approximated by a simpler function
$G_a^{(2+)}$,
\begin{align}\label{17}
    G_a^{(2+)}(E,\bfx_1,&\,\bfx_2)  =
    G^{(0)}(E+\Omega,\bfx_1,\bfx_2)
  \nonumber \\ &
   - G^{(0)}(E,\bfx_1,\bfx_2)
   -         \Omega\, \frac{\partial}{\partial E}\,
             G^{(0)}(E,\bfx_1,\bfx_2)\,,
\end{align}
where
\begin{equation}\label{18}
    \Omega  = \frac{2\Za}{x_1+x_2}\,.
\end{equation}
The above approximation can be obtained from the exact expression
for $G^{(2+)}$ by neglecting the commutators $[V,G^{(0)}]$ to all
orders. The function $G^{(2+)}_a$ is expressed in terms of the
free Green function and can be easily evaluated in a closed form.
We thus write $\Delta E_{\rm \, many}$ as a sum of two terms,
\begin{equation}\label{19}
    \Delta E_{\rm \, many} = \Delta E_{\rm \, many}^{\rm
\, sub}+ \Delta E_{\rm \, many}^{\rm \, remd} \,.
\end{equation}
The subtraction term $\Delta E_{\rm \, many}^{\rm \, sub}$ is
obtained from the high-energy part of Eq.~(\ref{16}) by the
substitution $G^{(2+)}\to G^{(2+)}_a$. The second term $\Delta
E_{\rm \, many}^{\rm \, remd}$ is the remainder. The subtraction
term is evaluated numerically in its closed form ({\em i.e.,}
without any partial-wave expansion), whereas the remainder yields
a rapidly converging partial-wave expansion; for the details of
the evaluation see Ref.~\cite{yerokhin:05:se}.

%%%%%%%%%%%%%%%%%%%%%%%%%%%%%%%%%%%%%%%%%%%%%%%
\subsection{Reducible part}

The reducible part is defined by Eq.~(\ref{8}). Using the
definition of the self-energy function and employing the contour
$C_{LH}$ for the integration over $\omega$, we write the
expression in the form
\begin{align} \label{20}
\Delta E_{\rm red} =&\ 2\,\im\alpha\, \delta \vare_a
\int_{C_{LH}} d\omega\,\int d\bfx_1\,d\bfx_2\,
    D^{\mu\nu}(\omega,\bfx_{12})\,
 \nonumber \\ & \times
  \psi_a^{\dag}(\bfx_1)
      \alpha_{\mu}\,
      \left. \frac{\partial}{\partial \vare}\,
         G(\vare-\omega,\bfx_1,\bfx_2)
    \right|_{\vare = \vare_a}\!\!\!\!
         \alpha_{\nu}\,
         \psi_a(\bfx_2)\,.
         \nonumber \\
\end{align}
Ultraviolet (UV) and infrared (IR) divergences present in this
expression can be conveniently isolated by separating the Green
function $G$ into 3 parts,
\begin{equation}\label{21}
    G(E) = G^{(0)}(E)+ G^{(a)}(E)+ \bigl[G(E) - G^{(0)}(E)- G^{(a)}(E)\bigr]\,,
\end{equation}
where $G^{(a)}$ incorporates the part of the spectral
decomposition of the bound-electron propagator with $\vare_n =
\vare_a$,
\begin{equation}\label{22}
    G^{(a)}(E,\bfx_1,\bfx_2) = \sum_{\mu_a}
        \frac{\psi_a(\bfx_1)\,\psi_a^{\dag}(\bfx_2)}{E-\vare_a(1-\im 0)}\,,
\end{equation}
and $\mu_a$ denotes the momentum projection of the states $\psi_a$
in this expression. The terms on the right-hand-side of
Eq.~(\ref{21}) substituted in Eq.~(\ref{20}) give rise to the
splitting of $\Delta E_{\rm ir}$, respectively, into 3 parts:
\begin{equation}\label{23}
    \Delta E_{\rm red} = \Delta E_{\rm red}^{(0)}+ \Delta E_{\rm red}^{(a)}
       + \Delta E_{\rm red}^{\rm \, many}\,.
\end{equation}
In this sum, the term $\Delta E_{\rm red}^{(0)}$ contains all UV
divergences. It is calculated in momentum space in a way similar
to that for the zero-potential part of the first-order self-energy
correction. UV-divergent terms are covariantly isolated; they
disappear when combined with the free-propagator contribution in
the vertex part. The term $\Delta E_{\rm red}^{(a)}$ contains all
IR divergences present in the reducible part. (IR divergences of
this type are sometimes also termed as the {\em reference-state}
singularities). These divergences are regularized by employing the
photon propagator with a finite photon mass $\mu$. The limit
$\mu\to 0$ can be taken when $\Delta E_{\rm red}^{(a)}$ is
combined with the corresponding contribution from the vertex part.

The term $\Delta E_{\rm red}^{\rm \, many}$ is finite.
Contributions of such type are frequently encountered in all-order
QED calculations. Usually, they are evaluated in coordinate space
after expanding into an infinite partial-wave series. In the
present investigation, we modify the standard scheme in order to
achieve a better convergence of the partial-wave expansion,
analogously to that for  the irreducible part. We thus separate
$\Delta E_{\rm red}^{\rm \, many}$ into two parts,
\begin{equation}\label{24}
    \Delta E_{\rm red}^{\rm \, many} = \Delta E_{\rm red}^{\rm \, sub}
       + \Delta E_{\rm red}^{\rm \, remd}\,.
\end{equation}
The remainder term $\Delta E_{\rm red}^{\rm \, remd}$ is obtained
from $\Delta E_{\rm red}^{\rm \, many}$ by replacing the standard
subtraction $\bigl[G(E) - G^{(0)}(E)- G^{(a)}(E)\bigr]$ by
$\bigl[G(E) - G^{(0)}(E+\Omega)- G^{(a)}(E)\bigr]$ in the
high-energy part of the expression. The remaining difference
$\bigl[G^{(0)}(E+\Omega)- G^{(0)}(E)\bigr]$ gives rise to the
subtraction term $\Delta E_{\rm red}^{\rm \, sub}$. More
explicitly, the subtraction term is written as
\begin{eqnarray} \label{25}
\Delta E_{\rm red}^{\rm \, sub} &=&2\,\im\alpha\, \delta \vare_a
\int_{C_{H}} d\omega\,\int d\bfx_1\,d\bfx_2\,
    D^{\mu\nu}(\omega,\bfx_{12})\,
 \nonumber \\ && \times
\psi_a^{\dag}(\bfx_1)\,
      \alpha_{\mu}\,
   \frac{\partial}{\partial \vare}\,
         \Bigl[ G^{(0)}(\vare-\omega+\Omega,\bfx_1,\bfx_2)
 \nonumber \\ &&
    \left.
           - G^{(0)}(\vare-\omega,\bfx_1,\bfx_2) \Bigr]
    \right|_{\vare = \vare_a}\!\!\!\!
         \alpha_{\nu}\,
         \psi_a(\bfx_2)\,,
\end{eqnarray}
where $\Omega$ is given by Eq.~(\ref{18}). This expression is
calculated in its closed form in coordinate space. The calculational
formulas are immediately obtained from the corresponding
expressions for the subtraction term for the first-order
self-energy correction \cite{yerokhin:05:se}.
The remainder term is calculated by a partial-wave expansion. Due to
the additional subtraction in the high-energy part, the convergence
properties of this partial-wave expansion are much better than in
the standard approach.

%%%%%%%%%%%%%%%%%%%%%%%%%%%%%%%%%%%%%%%%%%%%%%%
\subsection{Vertex part}

Rewriting expression~(\ref{9}) for the vertex part of the
self-energy \hfs correction in terms of the bound-electron
propagators, we obtain
\begin{eqnarray}   \label{26}
\Delta E_{\rm ver} &=& 2\,\im\alpha \int_{C_{LH}} d\omega\,
           \int d\bm{x}_1\,d\bm{x}_2\,d\bm{x}_3\,
          \psi_a^{\dag}(\bm{x}_1)\, \alpha_{\mu}\,
  \nonumber \\ && \times
         G(\vare_a-\omega,\bm{x}_1,\bm{x}_2)\,
       \delta V(\bm{x}_2)\,
         G(\vare_a-\omega,\bm{x}_2,\bm{x}_3)\,
  \nonumber \\ && \times
         \alpha_{\nu}\,
          \psi_a(\bm{x}_3)\, D^{\mu\nu}(\omega,\bm{x}_{13})\,.
\end{eqnarray}
UV and IR divergences present in this expression
can be conveniently isolated by the following separation
\begin{eqnarray}    \label{27}
G\,\delta V\,G &=& G^{(0)}\,\delta V\,G^{(0)} +
              G^{(a)}\,\delta V\,G^{(a)}
\nonumber \\ &&
              +  \left[ G\,\delta V\,G - G^{(0)}\,\delta V\,G^{(0)} - G^{(a)}\,\delta V\,G^{(a)}\right]\,.
\nonumber \\
\end{eqnarray}
This separation, being substituted into Eq.~(\ref{26}), gives rise
to the following three parts of $\Delta E_{\rm ver}$,
respectively,
\begin{equation}\label{28}
    \Delta E_{\rm ver} = \Delta E_{\rm ver}^{(0)}+ \Delta E_{\rm ver}^{(a)}
       + \Delta E_{\rm ver}^{\rm \, many}\,.
\end{equation}
Only the first term in this sum is UV divergent. UV divergences in
$\Delta E_{\rm ver}^{(0)}$ are covariantly isolated by employing a
momentum-space representation; they disappear when combined with
the corresponding contribution from the reducible part, see, {\em
e.g.}, Ref.~\cite{blundell:97:pra}. The second term $\Delta E_{\rm
ver}^{(a)}$ is IR divergent. In order to retain its finite part,
we consider it together with the corresponding contribution from
the reducible part,
\begin{eqnarray}   \label{29}
\Delta E_{\rm ver}^{(a)} + \Delta E_{\rm red}^{(a)} &=&
    \frac{\im}{2\pi} \int_{C_{LH}}d\omega\,
      \frac{1}{(\omega-\im 0)^2}\,
  \nonumber \\ && \times
         \sum_{\mu_{a'}\mu_{a''}}\,
           \biggl[ \lbr a'|\delta V|a''\rbr\,
                        \lbr aa''|I(\omega)|a'a\rbr
  \nonumber \\ &&
   - \lbr a|\delta V|a\rbr\,
                        \lbr aa'|I(\omega)|a'a\rbr
       \biggr]\,,
\end{eqnarray}
where $a'$ and $a''$ denote the intermediate states with
$\vare_n=\vare_a$ and with the momentum projection $\mu_{a'}$ and
$\mu_{a''}$, respectively. The integration over $\omega$ can be
carried out analytically, which leads to an explicitly finite
result. Sometimes it is more convenient to calculate this
contribution directly according to Eq.~(\ref{29}) (as long as the
contour $C_{LH}$ is employed for the integration over $\omega$,
this expression is suitable for the numerical evaluation).

The third term $\Delta E_{\rm ver}^{\rm \, many}$ does not contain
any divergences and is calculated in coordinate space after
expanding into a partial-wave series. We note that the additional
subtraction, similar to the one introduced for the reducible part,
does not improve the convergence properties of the partial-wave
expansion in this case. This is due to the fact that a significant
contribution to the partial-wave expansion terms originates from
the first-order commutator $[V,G^{(0)}]$. In order to achieve a
significant improvement, one needs to separate the complete
contribution of the vertex with one magnetic and one Coulomb
interaction. Such contribution was evaluated in a closed form for
the self-energy correction to the $g$-factor
\cite{yerokhin:02:prl,yerokhin:04:pra} using the explicit form of
the interaction with the constant magnetic field. In the case of
the self-energy \hfs correction, we are presently unable to obtain
a closed representation for this term. Nevertheless, the
partial-wave expansion for $\Delta E_{\rm ver}^{\rm \, many}$ is
converging significantly faster than that for $\Delta E_{\rm
ir}^{\rm \, many}$ and $\Delta E_{\rm red}^{\rm \, many}$, so that
the enhanced convergence achieved for the irreducible and
reducible parts results finally in a significant improvement of
the total accuracy of the calculation.

%%%%%%%%%%%%%%%%%%%%%%%%%%%%%%%%%%%%%%%%%%%%%%%
\subsection{Numerical results}
\label{sec:se:res}

The self-energy \hfs correction is conveniently represented in
terms of the dimensionless function $D_n^{\rm \, SE}$ defined as
\begin{equation}\label{2}
    \Delta E_{\rm SE} = \frac{E_F}{n^3}\, \frac{\alpha}{\pi}\, D_n^{\rm \, SE}(\Za)\,,
\end{equation}
where $n$ is the principal quantum number. The results of our
numerical calculation for the individual contributions of this
correction for the $1s$ and $2s$ states and $Z=10$ are presented
in Table~\ref{tab:breakdown}. The calculation was performed for
the point nuclear model and in the Feynman gauge. In
Table~\ref{tab:total} we list the final results for the
self-energy \hfs correction for H-like ions with $Z$ varying from
5 to 30. A comparison of the results of different theoretical
evaluations for this correction in the low-$Z$ region is given in
Table~\ref{tab:compar}.

%%%%%%%%%%%%%%%%%%%%%%%%%%%%%%%%%%%%%%%%%%%%%%%%%%%%%%%%%%%%%%%%%%%%
%
% Table I
%
%%%%%%%%%%%%%%%%%%%%%%%%%%%%%%%%%%%%%%%%%%%%%%%%%%%%%%%%%%%%%%%%%%%%
\begingroup
\squeezetable
\begin{table*}
\begin{center}
\begin{minipage}{16.0cm}
\caption{ Individual contributions to the self-energy \hfs
correction for $Z=10$, in units of the function $D_n$ defined by
Eq.~(\ref{3}). \label{tab:breakdown} }
\begin{ruledtabular}
\begin{tabular}{ccccccc}
     &  $\Delta E_{\rm ir}$ &  $\Delta E_{\rm red}^{(0)}+\Delta E_{\rm ver}^{(0)}$
                                   & $\Delta E_{\rm red}^{(a)}+\Delta E_{\rm ver}^{(a)}$
                                           & $\Delta E_{\rm red}^{\rm \, many}$
                                                             & $\Delta E_{\rm ver}^{\rm \, many}$
                                                                 & Total
\\[0.5pt]
\hline
1$s$ & $-$0.263902 &  1.896440  &$-$0.002385 & $-$1.607827 & $-$0.185180& $-$0.162853(1) \\
2$s$ & $-$0.223649 &  3.433554  &$\ \ \ $0.176267 & $-$1.878883 & $-$1.649848& $-$0.142559(3) \\
\end{tabular}
\end{ruledtabular}
\end{minipage}
\end{center}
\end{table*}
\endgroup

%%%%%%%%%%%%%%%%%%%%%%%%%%%%%%%%%%%%%%%%%%%%%%%%%%%%%%%%%%%%%%%%%%%%
%
% Table II
%
%%%%%%%%%%%%%%%%%%%%%%%%%%%%%%%%%%%%%%%%%%%%%%%%%%%%%%%%%%%%%%%%%%%%
\begin{table}
\caption{The self-energy \hfs correction for the $n=1$ and $n=2$
states of light H-like ions. \label{tab:total}}
\begin{ruledtabular}
\begin{tabular}{r..}
 $Z$ & \multicolumn{1}{c}{$D_1(\Za)$}& \multicolumn{1}{c}{$D_2(\Za)$}\\
\hline
5    &    0.174x\,026\,(2)   &     0.181x\,940\,(2) \\
6    &    0.106x\,815\,(2)   &     0.117x\,124\,(2) \\
7    &    0.039x\,476\,(2)   &     0.052x\,265\,(2) \\
8    &   -0.027x\,933\,(1)   &    -0.012x\,626\,(2) \\
9    &   -0.095x\,379\,(1)   &    -0.077x\,559\,(3) \\
10   &   -0.162x\,853\,(1)   &    -0.142x\,559\,(3) \\
12   &   -0.297x\,905\,(1)   &    -0.272x\,913\,(1) \\
15   &   -0.501x\,056\,(1)   &    -0.470x\,078\,(1) \\
20   &   -0.843x\,572\,(1)   &    -0.807x\,153\,(1) \\
25   &   -1.196x\,242\,(2)   &    -1.162x\,717\,(2) \\
30   &   -1.566x\,491\,(3)   &    -1.547x\,535\,(2) \\
\end{tabular}
\end{ruledtabular}
\end{table}

%%%%%%%%%%%%%%%%%%%%%%%%%%%%%%%%%%%%%%%%%%%%%%%%%%%%%%%%%%%%%%%%%%%%
%
% Table III
%
%%%%%%%%%%%%%%%%%%%%%%%%%%%%%%%%%%%%%%%%%%%%%%%%%%%%%%%%%%%%%%%%%%%%
\begin{table}
%\begin{center}
%\begin{minipage}{16.0cm}
\caption{ Comparison of the results of different calculations of
the self-energy \hfs correction for the $1s$ and $2s$ states in
light H-like ions, in units of $D_n(\Za)$. \label{tab:compar} }
\begin{ruledtabular}
\begin{tabular}{c...c}
           &  \multicolumn{1}{c}{$Z=5$}&   \multicolumn{1}{c}{$Z=10$}& \multicolumn{1}{c}{$Z=20$}  & Ref.
\\ \hline
$1s$       &  0.174x\,026\,(2)         &   -0.162x\,853\,(1)     &  -0.843x\,572\,(1)   &     \\
           &  0.174x\,028\,(20)        &   -0.162x\,860\,(20)    &  -0.843x\,588\,(15)  & \cite{yerokhin:01:hfs}\\
           &  0.174x\,05\,(1)          &   -0.162x\,83\,(1)      &  -0.843x\,56\,(1)    & \cite{blundell:97:prl}\\
           &  0.174x\,1(1)             &   -0.162x\,8(1)         &                      & \cite{sunnergren:98:pra}\\
\hline
$2s$       &  0.181x\,940\,(2)         &   -0.142x\,559\,(3)     &  -0.807x\,153\,(1)   &    \\
           &  0.181x\,96\,(10)         &   -0.142x\,51\,(10)     &  -0.807x\,16\,(6)    & \cite{yerokhin:01:hfs}\\
\end{tabular}
\end{ruledtabular}
%\end{minipage}
%\end{center}
\end{table}

\begin{figure}
\includegraphics[width=0.95\columnwidth]{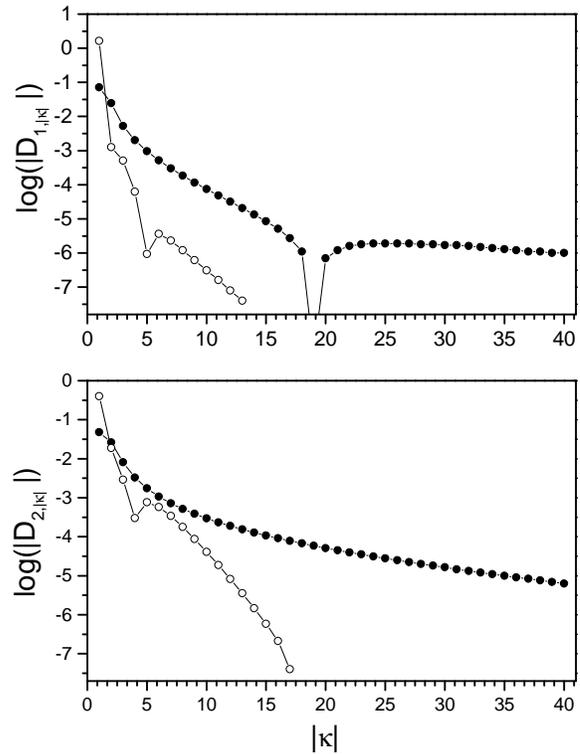}
\caption{\label{fig1} The absolute magnitude of the individual
terms of the partial-wave expansion for the irreducible part of
the self-energy \hfs  correction for $Z=5$, within the standard
potential-expansion scheme (filled dots) and with the additional
subtraction employed in the present work (open dots). Plotted are
the contributions to the function $D_n(\Za)$ as a function of the
absolute value of the relativistic angular momentum parameter
$\kappa$. A discontinuity of the curve on the upper graph around
$|\kappa|=19$ is due to the change of the sign of the
contributions to $D_1(\Za)$. }
\end{figure}

\begin{figure}
\includegraphics[width=0.95\columnwidth]{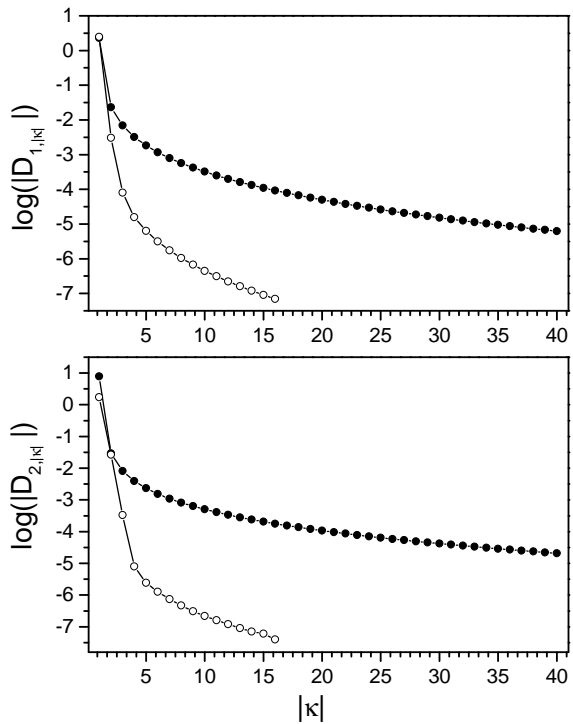}
\caption{\label{fig2} The same as in Fig.~\ref{fig1}, but for the
reducible part of the self-energy \hfs correction. }
\end{figure}

As can be seen from Table~\ref{tab:compar}, the present
calculation improves the numerical accuracy of the self-energy
\hfs correction by about an order of magnitude for the $1s$ state
and even more for the $2s$ state, as compared to our previous
calculation \cite{yerokhin:01:hfs}. This progress is due to the
additional subtraction scheme employed in the present work for the
evaluation of the irreducible and reducible parts of the
correction. In order to illustrate the improvement in the
convergence properties of the partial-wave expansion introduced by
this scheme, in Figs.~\ref{fig1} and \ref{fig2} we plot (in the
decimal logarithmic scale) the dependence of the absolute value of
the individual terms of the partial-wave series on the expansion
parameter $|\kappa|$ within the standard potential-expansion
approach and within the new subtraction scheme. (The parameter
$\kappa$ is the Dirac angular-momentum eigenvalue of one of the
electron propagators in the vertex function.) Figs.~\ref{fig1} and
\ref{fig2} represent this comparison for the irreducible and the
reducible part, respectively.

%%%%%%%%%%%%%%%%%%%%%%%%%%%%%%%%%%%%%%%%%%%%%%%%%%%%%%%%%%%%%%%%%%%%
%
% Figure I
%
%%%%%%%%%%%%%%%%%%%%%%%%%%%%%%%%%%%%%%%%%%%%%%%%%%%%%%%%%%%%%%%%%%%%

%%%%%%%%%%%%%%%%%%%%%%%%%%%%%%%%%%%%%%%%%%%%%%%%%%%%%%%%%%%%%%%%%%%%
%
% Figure II
%
%%%%%%%%%%%%%%%%%%%%%%%%%%%%%%%%%%%%%%%%%%%%%%%%%%%%%%%%%%%%%%%%%%%%

As a result of the improvement achieved, we were able to eliminate
completely the uncertainty arising from termination of the
partial-wave expansion in the irreducible and reducible parts.
Still, there remains the partial-wave expansion of the
many-potential vertex term $\Delta E_{\rm ver}^{\rm \, many}$,
which has to be terminated and properly extrapolated to infinity.
(In actual calculations, the summation was terminated at
$|\kappa|=40$). The error due to this extrapolation yields one of
the main uncertainties of our numerical evaluation (the other
source of the uncertainty is the stability of numerical
integrations.) Fortunately, the partial-wave expansion of $\Delta
E_{\rm ver}^{\rm \, many}$ is monotonic and relatively
well-converging (not worse than $1/|\kappa|^3$ for all $Z\ge 5$),
and so the uncalculated tail of the expansion can be estimated
reasonably well.

%%%%%%%%%%%%%%%%%%%%%%%%%%%%%%%%%%%%%%%%%%%%%%%
%
%
%
%%%%%%%%%%%%%%%%%%%%%%%%%%%%%%%%%%%%%%%%%%%%%%%
\section{Vacuum-polarization correction}
\label{sec:vp}

The vacuum-polarization \hfs correction can be conveniently split
into two parts, the so-called {\em electric-} and {\em
magnetic-loop} contributions.  The electric-loop part originates
from the diagrams with the \hfs interaction attached to the
external electron line, whereas the magnetic-loop one comes from
the diagram with the \hfs interaction attached to the
vacuum-polarization loop. These contributions are also
traditionally separated into the Uehling and Wichmann-Kroll (WK)
parts. The WK part is suppressed by a factor of $(\Za)^2$ as
compared to the Uehling contribution and can often be regarded as
a small correction for low-$Z$ ions.

The $\Za$ expansion of the one-loop vacuum-polarization \hfs
correction reads
\begin{eqnarray}
  D_n^{\rm VP}(\Za) &=& (\Za)\,b_{10}+ (\Za)^2\Bigl[
     L\,b_{21}+b_{20}\Bigr]
   \nonumber \\ &&
     + (\Za)^3\Bigl[L\,b_{31}+ F_{n}^{\rm VP}(\Za)\Bigr]\,,
\end{eqnarray}
where $L =\ln[(Z\alpha)^{-2}]$, the function $D_n^{\rm VP}$ is
related to the energy shift analogously to Eq.~(\ref{2}), and the
function $F_{n}^{\rm VP}$ incorporates all higher-order terms,
$F_{n}^{\rm VP}(\Za) = b_{30}+ \Za \,(\ldots)\,$. The first
expansion coefficients up to $b_{31}$ stem from the Uehling part
of the vacuum-polarization correction; they are given by (see,
{\em e.g.}, \cite{karshenboim:02:epjd})
\begin{eqnarray}
  b_{10}(ns) &=&  \frac{3\,\pi}{4}\,, \\
  b_{21}(ns) &=& \frac{8}{30}\,,  \\
  b_{20}(1s) &=& \frac{34}{225}-\frac{8}{15}\,\ln 2\,,  \\
  b_{20}(2s) &=& -\frac{247}{450}\,, \\
  b_{31}(ns) &=& \frac{13\,\pi}{48}\,.
\end{eqnarray}
Higher-order coefficients starting with $b_{30}$ arise both from
the Uehling and WK contributions; only their Uehling part is
presently known \cite{karshenboim:01:jetp,karshenboim:02:epjd}.

The Uehling part of the one-loop vacuum-polarization \hfs
correction can easily be calculated numerically; some results can
be found, {\em e.g.}, in
Refs.~\cite{sunnergren:98:pra,artemyev:01}. In the case of the
point nucleus, this contribution was evaluated also analytically
\cite{karshenboim:01:jetp,karshenboim:02:epjd}. For completeness,
we re-calculate it in the present work. The corresponding
contributions to the higher-order remainder $F_{n}^{\rm VP}(\Za)$
for the point nuclear model are listed in the first and the second
columns of Table~~\ref{tab:vp} for the $1s$ and $2s$ states,
respectively. The results presented are in agreement with the
previous calculations of this correction.

%%%%%%%%%%%%%%%%%%%%%%%%%%%%%%%%%%%%%%%%%%%%%%%%%%%%%%%%%%%%%%%%%%%%
%
% Table IV
%
%%%%%%%%%%%%%%%%%%%%%%%%%%%%%%%%%%%%%%%%%%%%%%%%%%%%%%%%%%%%%%%%%%%%
\begin{table*}
\begin{center}
\begin{minipage}{16.0cm}
\caption{Individual contributions to $F^{\rm VP}_n(\Za)$ for the
$1s$ and $2s$ states of light H-like ions. \label{tab:vp}}
\begin{ruledtabular}
\begin{tabular}{r......}
 $Z$ &  \multicolumn{2}{c}{Uehling}&
         \multicolumn{2}{c}{Electric-loop WK}& \multicolumn{2}{c}{Magnetic-loop WK}\\
&  \multicolumn{1}{c}{$1s$} & \multicolumn{1}{c}{$2s$}     &  \multicolumn{1}{c}{$1s$} & \multicolumn{1}{c}{$2s$} & \multicolumn{1}{c}{$1s$} & \multicolumn{1}{c}{$2s$} \\
\hline
1    &   7.x231  &   9.x546  &  -0.x117   &  -0.x117  &              &            \\
2    &   7.x337  &   9.x651  &  -0.x120   &  -0.x119  &              &            \\
5    &   7.x587  &   9.x901  &  -0.x128   &  -0.x125  &              &            \\
10   &   7.x947  &  10.x282  &  -0.x138   &  -0.x133  &  -0.x699(2)  &  -0.x706(2) \\
     &     x     &     x     &  -0.x139\,^a &           &  -0.x697\,^a   &            \\
12   &   8.x092  &  10.x441  &  -0.x142   &  -0.x136  &  -0.x701(2)  &  -0.x709(2) \\
14   &   8.x240  &  10.x609  &  -0.x145   &  -0.x138  &  -0.x703(2)  &  -0.x714(2) \\
16   &   8.x394  &  10.x788  &  -0.x149   &  -0.x141  &  -0.x705(2)  &  -0.x718(2) \\
18   &   8.x556  &  10.x978  &  -0.x153   &  -0.x145  &  -0.x707(2)  &  -0.x723(2) \\
     &     x     &     x     &  -0.x154\,^a &           &  -0.x706\,^a   &            \\
20   &   8.x725  &  11.x182  &  -0.x156   &  -0.x148  &  -0.x712(2)  &  -0.x732(2) \\
22   &   8.x904  &  11.x401  &  -0.x160   &  -0.x151  &  -0.x717(2)  &  -0.x740(2) \\
24   &   9.x093  &  11.x635  &  -0.x164   &  -0.x155  &  -0.x725(2)  &  -0.x752(2) \\
\end{tabular}

$^a$ Ref.~\cite{sunnergren:98:pra}\,.
\end{ruledtabular}
\end{minipage}
\end{center}
\end{table*}

The WK part of the vacuum-polarization \hfs correction is more
difficult to calculate. Especially, this refers to the
magnetic-loop WK contribution. As outlined in
Ref.~\cite{artemyev:01}, this correction is divergent in the
point-dipole approximation for the nuclear magnetization
distribution. A finite result for this correction is obtained if
an extended nuclear magnetization distribution is employed. It
should be also taken into account that the magnetic-loop WK
interaction contributes to the measured value of the nuclear
magnetic moment \cite{milstein:89}. In order to prevent
double-counting, the corresponding contribution should be
subtracted from the magnetic-loop WK part of the
vacuum-polarization \hfs correction. Practical calculations
\cite{sunnergren:98:pra,artemyev:01} show that, after such a
subtraction, the magnetic-loop WK correction depends weakly on
details of the nuclear magnetization distribution and has a finite
limit in the point-dipole approximation.

The results of our numerical evaluation of the electric- and
magnetic-loop WK contributions for the $1s$ and $2s$ states of
light H-like ions are listed in Table~\ref{tab:vp}, in terms of
the higher-order remainder $F_{n}^{\rm VP}(\Za)$. The
electric-loop WK correction was calculated for the point nuclear
model by employing the analytical-approximation formulas for the
WK potential from Ref.~\cite{fainshtein:91}. The relative accuracy
of this approximation is considered by the authors to be not worse
than $10^{-4}$ for all $Z$ up to $\Za=0.95$. As an independent
test of the accuracy of this approximation in the low-$Z$ region,
we checked that it reproduces well the first two terms of the
$\Za$ expansion of the WK correction to the Lamb shift. The
magnetic-loop WK correction was calculated for the point-dipole
nuclear magnetization model by using a code developed in our
previous investigation \cite{artemyev:01}. A comparison given in
Table~\ref{tab:vp} demonstrates good agreement of our numerical
values with the $1s$ results of Ref.~\cite{sunnergren:98:pra} for
$Z=10$ and 18.

%%%%%%%%%%%%%%%%%%%%%%%%%%%%%%%%%%%%%%%%%%%%%%%
%
%
%
%%%%%%%%%%%%%%%%%%%%%%%%%%%%%%%%%%%%%%%%%%%%%%%
\section{Higher-order one-loop QED correction}
\label{sec:ho}

One of the main goals of our investigation is to improve the
accuracy of the one-loop QED correction for $Z=1$ and 2, these
being the most interesting cases from the experimental point of
view. The present approach does not employ the $\Za$ expansion
and, therefore, our numerical results do not suffer from omission
of the higher-order terms, as is the case with the perturbative
$\Za$-expansion approach. But on the other hand, technical
problems do not presently allow us to perform a direct numerical
evaluation for $Z=1$ and 2 with a sufficient accuracy. In the
present work, we employ an indirect method used previously in
Refs.~\cite{blundell:97:prl,yerokhin:01:hfs}. By subtracting the
known terms of the $\Za$ expansion from the all-order results for
higher values of $Z$, we identify the higher-order remainder and
then extrapolate it to $Z=1$ and 2.

First we summarize the results obtained for the one-loop
self-energy \hfs correction within the perturbative
$\Za$-expansion approach. The corresponding $\Za$ expansion reads
\begin{align}\label{3}
  D_n^{\rm \, SE}(\Za) =&\ a_{00}+ (\Za)\,a_{10}+ (\Za)^2\Bigl[
     L^2 a_{22}+ L\,a_{21}+a_{20}\Bigr]
   \nonumber \\ &
     + (\Za)^3\Bigl[L\,a_{31}+ F_{n}^{\rm \, SE}(\Za)\Bigr]\,,
\end{align}
where  $L =\ln[(Z\alpha)^{-2}]$  and $F_{n}^{\rm \, SE}$ is the
remainder containing all higher-order terms, $F_{n}^{\rm \,
SE}(\Za) = a_{30}+ \Za \,(\ldots)\,$. The results presently
available for the expansion coefficients are (for the references
see, {\em e.g.}, \cite{karshenboim:02:epjd}):
\begin{eqnarray}\label{4}
  a_{00}(ns) &=& \frac12 \,, \\
  a_{10}(ns) &=& \left(\ln 2-\frac{13}{4}\right)\,\pi \,, \\
  a_{22}(ns) &=& -\frac23\,,\\
  a_{21}(1s) &=& -\frac{8}{3}\ln 2+\frac{37}{72}\,, \\
  a_{21}(2s) &=& a_{21}(1s)-\frac{8}{3}\ln 2+\frac{7}{2}\,, \\
  a_{20}(1s) &=& 17.122\,339\ldots \,, \\
  a_{20}(2s) &=& a_{20}(1s)-5.221\,233(3) \,,\\
  a_{31}(ns) &=& \left(\frac52 \ln 2-\frac{191}{32}\right)\, \pi\,.
\end{eqnarray}
For the $a_{30}$ term, there is a preliminary result \cite{nio:01}
for the $1s$ state, $a_{30}(1s) = -15.9(1.6)$, and a partial
result \cite{karshenboim:01:hfs,karshenboim:02:epjd} for the
difference $\Delta_{21}$, $a_{30}(2s)-a_{30}(1s) = 7.92$.

\begin{figure}
\includegraphics[width=\columnwidth]{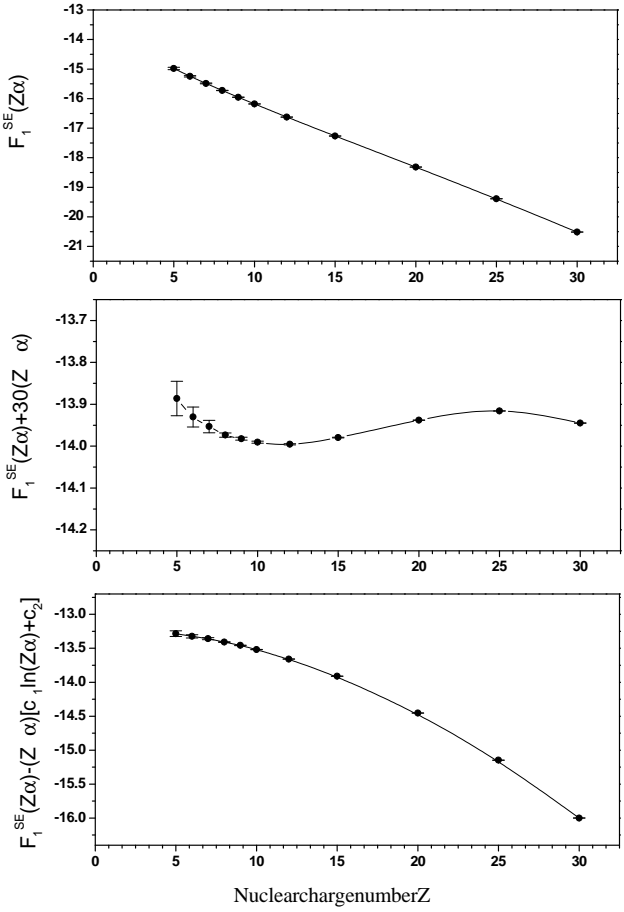}
\caption{\label{fig:fho1s} The higher-order part of the $1s$
self-energy \hfs correction $F_{1}^{\rm \, SE}$ as a function of
the nuclear charge number (the upper graph); $F_{1}^{\rm \, SE}$
with its linear part (obtained by a global linear fit) subtracted
(the middle graph); $F_{1}^{\rm \, SE}$ with its next-to-leading
contribution (obtained by a least-squares fit) subtracted (the
lower graph). The numerical values of the coefficients $c_1$ and
$c_2$ are: $c_1 = 14.83$ and $c_2 = 2.08$. }
\end{figure}

\begin{figure}
\includegraphics[width=\columnwidth]{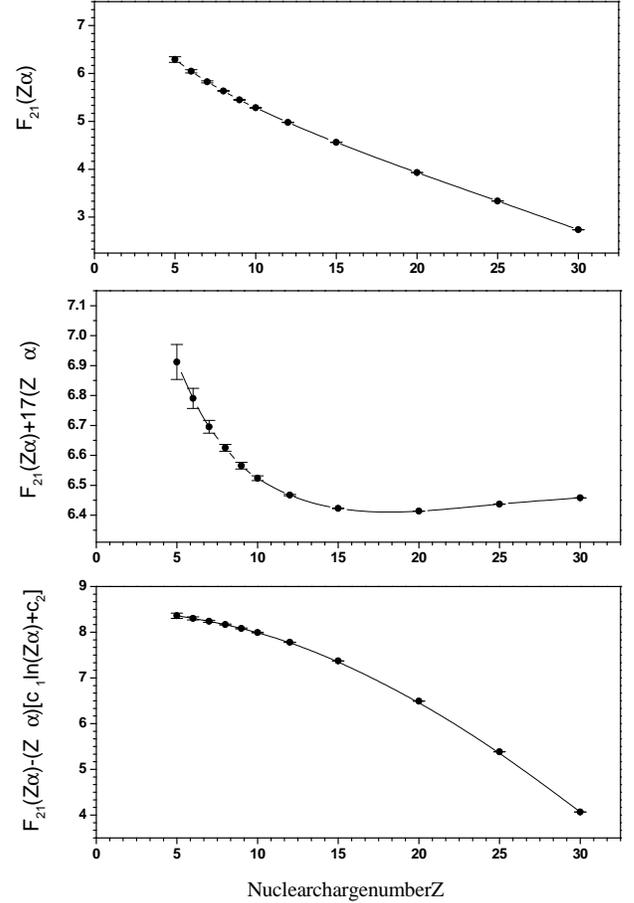}
\caption{\label{fig:fho2s1s} The same as in Fig.~\ref{fig:fho1s}
but for the difference $F_{21}^{\rm \, SE} = F_{2}^{\rm \,
SE}-F_{1}^{\rm \, SE}$. The numerical values of the coefficients
$c_1$ and $c_2$ are: $c_1 = 28.26$ and $c_2 = 36.85$. }
\end{figure}

The higher-order self-energy remainder $F_n^{\rm \, SE}$ can be
inferred from our all-order numerical data. The corresponding
results for the function $F_1^{\rm \, SE}(\Za)$ and the difference
$F_{21}^{\rm \, SE}(\Za)\equiv F_2^{\rm \, SE}(\Za)-F_1^{\rm \,
SE}(\Za)$ are plotted on the upper graphs of Figs.~\ref{fig:fho1s}
and \ref{fig:fho2s1s}, respectively. We note that both the
$F_1^{\rm \, SE}(\Za)$ and $F_{21}^{\rm \, SE}(\Za)$ functions
have a rapidly varying structure in the low-$Z$ region. In order
to demonstrate this more clearly, we subtract their ``linear''
part (obtained by a global linear fit), with the corresponding
plots presented on the middle graphs of Figs.~\ref{fig:fho1s} and
\ref{fig:fho2s1s}. The behavior observed apparently indicates that
the logarithmic term to order $\alpha(\Za)^4 E_F$ enters with a
large coefficient, which complicates extrapolation considerably.

%%%%%%%%%%%%%%%%%%%%%%%%%%%%%%%%%%%%%%%%%%%%%%%%%%%%%%%%%%%%%%%%%%%%
%
% Figure III
%
%%%%%%%%%%%%%%%%%%%%%%%%%%%%%%%%%%%%%%%%%%%%%%%%%%%%%%%%%%%%%%%%%%%%

%%%%%%%%%%%%%%%%%%%%%%%%%%%%%%%%%%%%%%%%%%%%%%%%%%%%%%%%%%%%%%%%%%%%
%
% Figure IV
%
%%%%%%%%%%%%%%%%%%%%%%%%%%%%%%%%%%%%%%%%%%%%%%%%%%%%%%%%%%%%%%%%%%%%

Now we would like to extrapolate our results for the higher-order
remainder to the lower values of $Z$, namely $Z=0$, 1, and 2. For
this purpose we employ a procedure similar to the one recently
described in detail in Ref.~\cite{bigot:03}. The extrapolated
value of a function at $Z= z_0$ is obtained in two steps. First we
apply an (exact) linear  fit to each two consecutive points from
our data set and store the resulting value at $Z=z_0$ as a
function of the average abscissa of the points involved in the
fit. Then we perform a global parabolic least-squares fit to the
set of data obtained and take the fitted value at $Z=z_0$ as a
final result.

We tested this procedure for variation of the logarithmic
contribution to the next-to-leading order and found it rather
stable. However, in order to take into account the presence of
such contribution explicitly, we modify the procedure described
above as following. First, we approximate our numerical data by a
function
\begin{equation}
f(Z) = c_0+ (\Za)\,[c_1\, \ln (\Za)+c_2+(\Za)\,c_3 ]\,
\end{equation}
with free coefficients $c_i$, which are determined by a
least-squares fit similar to the one described in
Ref.~\cite{ivanov:01}. Then, we use the values obtained for $c_1$
and $c_2$ in order to define a modified higher-order remainder
function as
\begin{equation}
\widetilde{F}_n^{\rm \, SE}(\Za) = F_n^{\rm \, SE}(\Za) -
(\Za)\,[c_1\,\ln (\Za)+c_2]\,.
\end{equation}
The numerical results for this function are plotted on the lower
graphs of Figs.~\ref{fig:fho1s} and \ref{fig:fho2s1s}. The
function $\widetilde{F}_n^{\rm \, SE}$ is much flatter in the
low-$Z$ region than $F_n^{\rm \, SE}$ and, therefore, is more
suitable for the extrapolation. We obtain our final results
extrapolating the function $\widetilde{F}_n^{\rm \, SE}$ by means
of the procedure described above. The numerical values of the
higher-order self-energy remainder obtained in this way are given
in the first line of Table~\ref{tab:qedho}. In the next 3 lines of
the table, we present the results of the previous evaluations
\cite{blundell:97:prl,yerokhin:01:hfs,nio:01}. The numerical
values obtained for the function $F_{21}^{\rm \, SE}$ in this work
fall slightly outside the error bars ascribed to our previous
results \cite{yerokhin:01:hfs}, as a consequence of the
logarithmic contribution to the next-to-leading order being
apparently much larger than it was assumed in our former work. Our
present values for the function $F_1^{\rm \, SE}$ are in a
marginal agreement with the result by Blundell {\em et al.}
\cite{blundell:97:prl} and deviate by 1.5$\,\sigma$ from the
preliminary result by Nio \cite{nio:01}.

%%%%%%%%%%%%%%%%%%%%%%%%%%%%%%%%%%%%%%%%%%%%%%%%%%%%%%%%%%%%%%%%%%%%
%
% Table V
%
%%%%%%%%%%%%%%%%%%%%%%%%%%%%%%%%%%%%%%%%%%%%%%%%%%%%%%%%%%%%%%%%%%%%
\begin{table*}
\begin{center}
\begin{minipage}{16.0cm}
\caption{One-loop higher-order QED correction. Acronyms are:
``SE'' denotes the self-energy contribution, ``Ue'' -- the Uehling
part, ``WK-EL'' -- the electric-loop WK part, ``WK-ML'' -- the
magnetic-loop WK part. \label{tab:qedho}}
\begin{ruledtabular}
\begin{tabular}{l......c}
        & \multicolumn{1}{c}{$F_1(0\alpha)$}
              & \multicolumn{1}{c}{$F_1(1\alpha)$}
                     & \multicolumn{1}{c}{$F_1(2\alpha)$}
                         & \multicolumn{1}{c}{$F_{21}(0\alpha)$}
                                & \multicolumn{1}{c}{$F_{21}(1\alpha)$}
                                           & \multicolumn{1}{c}{$F_{21}(2\alpha)$}
   &  Reference \\
\hline
  SE    & -13.x2(4)   &  -13.x8(3)   &  -14.x1(3)   &   8.x4(5)   &   7.x6(4)   &  7.x2(3)&   \\
        &             &  -12.x0(2.0) &              &             &             &         & \cite{blundell:97:prl}\\
        &             &  -14.x3(1.1) &  -14.x5(7)   &             &   6.x5(8)   &  6.x3(6)& \cite{yerokhin:01:hfs}\\
        & -15.x9(1.6) &              &              &             &             &         & \cite{nio:01}\\
  Ue    &   7.x06     &    7.x23     &    7.x34     &   2.x32     &   2.x32     &  2.x31  &   \\
  WK-EL &  -0.x11     &   -0.x12     &   -0.x12     &   0.x00     &   0.x00     &  0.x00  &   \\
  WK-ML &  -0.x69(15) &   -0.x69(12) &   -0.x69(7)  &   0.x00     &   0.x00     &  0.x00  &   \\
 Total  &  -6.x9(4)   &   -7.x4(3)   &   -7.x6(3)   &  10.x7(5)   &   9.x9(4)   &  9.x5(3)&   \\
\end{tabular}
\end{ruledtabular}
\end{minipage}
\end{center}
\end{table*}

The Uehling part of the vacuum-polarization \hfs correction is
given in the next line of Table~\ref{tab:qedho}. The numerical
values are taken from Table~\ref{tab:vp} for $Z=1$ and 2 and from
Refs.~\cite{karshenboim:01:jetp,karshenboim:02:epjd} for $Z=0$.
The electric-loop WK correction for $Z=1$ and 2 was calculated
directly in Sec.~\ref{sec:vp}; the corresponding numerical value
for $Z=0$ was obtained by a simple extrapolation. Extrapolation
was also employed in order to obtain the results for the
magnetic-loop WK part of the vacuum-polarization correction
presented in the table. The error bars specified are obtained
under the assumption that the logarithmic contribution to the
next-to-leading order enters with a coefficient of about 2.

We now turn to the experimental consequences of our calculation.
As demonstrated in Ref.~\cite{karshenboim:02:epjd}, the
higher-order self-energy correction is one of the major sources of
uncertainty of the theoretical prediction for the normalized
difference of the \hfs intervals $\Delta_{21} = 8\,
\nu_{2s}-\nu_{1s}$ for the $^3{\rm He}^+$ ion. Our present
calculation changes the theoretical value of this correction by
$-0.056$~kHz (as compared to our former result
\cite{yerokhin:01:hfs}) and improves its accuracy by a factor of
2. The resulting value of the one-loop QED contribution that
incorporates all orders in $\Za$ starting with the constant term
to order $\alpha(\Za)^3E_F$ for the $^3{\rm He}^+$ ion is given in
the first line of Table~\ref{tab:D21}. In the next lines of the
table, we give the total theoretical value for the difference
$\Delta_{21}(^3{\rm He}^+)$ taken from
Ref.~\cite{karshenboim:02:epjd}, this value modified by the
present calculation, and the corresponding experimental result. As
can be seen from the table, our calculation increases the
deviation of the theoretical prediction from the experimental
value from $0.9\,\sigma$ to $1.6\,\sigma$.

%%%%%%%%%%%%%%%%%%%%%%%%%%%%%%%%%%%%%%%%%%%%%%%%%%%%%%%%%%%%%%%%%%%%
%
% Table VI
%
%%%%%%%%%%%%%%%%%%%%%%%%%%%%%%%%%%%%%%%%%%%%%%%%%%%%%%%%%%%%%%%%%%%%
\begin{table}
%\begin{center}
%\begin{minipage}{16.0cm}
\caption{Normalized difference of the \hfs intervals $\Delta_{21}
= 8\, \nu_{2s}-\nu_{1s}$ for the $^3{\rm He}^+$ ion, in kHz.
\label{tab:D21}}
\begin{ruledtabular}
\begin{tabular}{l.c}
Higher-order QED correction  &       -0x.594\,(19)     &                \\
$\Delta_{21}$, old theory    &  -1\,190x.068\,(64)     &  \cite{karshenboim:02:epjd} \\
$\Delta_{21}$, new theory    &  -1\,190x.124\,(55)     &                \\
$\Delta_{21}$, experiment    &  -1\,189x.979\,(71)     &  \cite{schluessler:69}+\cite{prior:77} \\
\end{tabular}
\end{ruledtabular}
%\end{minipage}
%\end{center}
\end{table}

It should be noted that our numerical results for all corrections
at $Z=1$, except for the one for the magnetic-loop WK
contribution, can be directly applied to the hyperfine splitting
in muonium. Our calculation of the magnetic-loop WK correction
does not hold for muonium since it involves a regularization by an
extended magnetization distribution of the nucleus and a
subtraction of the related contribution to the measured value of
the nuclear magnetic moment. In the case of muonium, the nucleus
is substituted by a point-like muon and the regularization should
be performed by a finite mass of the muon rather than by a finite
size.

%%%%%%%%%%%%%%%%%%%%%%%%%%
\section{Summary}

In the present investigation, we carried out an all-order (in
$\Za$) calculation of the one-loop QED correction to the hyperfine
splitting of the $1s$ and $2s$ states in light H-like ions. This
calculation significantly improved the accuracy of this
correction, as compared to the previous evaluations. By
subtracting the known terms of the $\Za$ expansion and
extrapolating the remainder to lower values of $Z$, we obtained
the results for the higher-order remainder for $Z=0$, 1, and 2.
Our calculation shifts the theoretical value of the normalized
difference $\Delta_{21}$ of the $1s$ and $2s$ \hfs intervals in
$^3{\rm He}^+$ by $0.056$~kHz and slightly improves its accuracy.

\section*{Acknowledgements}

This work was supported in part by RFBR (Grant No.~04-02-17574) and by
DFG.  A.N.A. and V.A.Y. acknowledge support from the ``Dynasty''
foundation and from INTAS (Grants No.~YS~03-55-960 and YS~03-55-1442).
G.P. acknowledges financial support from BMBF and GSI.

%%%%%%%%%%%%%%%%%%%%%%%%%%

%\bibliographystyle{d:/papers/bibtex/phaip30}
%\bibliography{d:/papers/bibtex/hfst}

\end{document}